\documentclass[aps, prb, superscriptaddress, twocolumn, letterpaper, 10pt, amsfonts, amsmath, amssymb, groupedaddress, longbibliography]{revtex4-2}
\usepackage{graphicx}

\usepackage{physics}
\usepackage{float}
\usepackage{xcolor}
\usepackage{soul}
\usepackage[normalem]{ulem}
\usepackage[mathscr]{euscript}
\usepackage{bm}
\usepackage{dsfont}
\usepackage[colorlinks=true,citecolor=blue,urlcolor=blue,linkcolor=blue]{hyperref}
\usepackage{float}

\newcommand{\ii}{\mathrm{i}}
\newcommand{\ee}{\mathrm{e}}
\renewcommand{\Vec}[1]{\bm{#1}}
\newcommand{\hc}{{\rm{H.c.}}}
\newcommand{\T}{{\rm{T}}}
\newcommand{\magnetic}{{\rm{mag}}}
\newcommand{\BZ}{{\rm{BZ}}} 
\newcommand{\B}{{\rm{B}}}
\newcommand{\skyrmion}{{\rm{skyrmion}}}
\newcommand{\conj}[1]{{#1}^*}
\newcommand{\skyr}{{$\mathbb{C}P^2$}\,}
\newcommand{\figref}{FIG.~\ref}
\newcommand{\secref}{Section \ref}
\newcommand{\appendixref}{Appendix \ref}
\renewcommand{\eqref}[1]{{Eq.~(\ref{#1})}}

\begin{document}
\title{Demonstration of \skyr skyrmions in three-band superconductors by self-consistent solutions to a Bogoliubov-de Gennes model}

\author{Andrea Benfenati} 
\affiliation{Department of Physics, The Royal Institute of Technology, Stockholm SE-10691, Sweden}
\author{Mats Barkman} 
\affiliation{Department of Physics, The Royal Institute of Technology, Stockholm SE-10691, Sweden}
\author{Egor Babaev} 
\affiliation{Department of Physics, The Royal Institute of Technology, Stockholm SE-10691, Sweden}
\begin{abstract}
Topological defects, such as magnetic-flux-carrying quantum vortices determine the magnetic response of superconductors and hence are of fundamental importance. Here, we show that stable $\mathbb{C}P^2$  skyrmions exist in three-band $s+\rm{i} s$ superconductors as fully self-consistent solutions to a microscopic Bogoluibov-de Gennes model. This allows us to calculate microscopically the magnetic signatures of $\mathbb{C}P^2$ skyrmions and their footprint in the local density of states.
\end{abstract}

\maketitle
\section{Introduction}

Skyrmions are topological solitons that  were originally discussed as
an effective description of nucleons \cite{skyrme1962unified}.
Since the first work, many generalizations were proposed \cite{Manton2004}.
Currently, the most studied case are magnetic skyrmions 
which is a topological defect characterized by so-called $S^2 \to S^2
$ topological map \cite{nagaosa2013topological}. However, more complex skyrmionic solutions exist in field theories with more components \cite{Manton2004}. In particular, early works  considered skyrmions in  $N$-component nonlinear  $\sigma$-models with a high broken symmetry \cite{golo1978solution,d19781n,Manton2004}.
More complicated skyrmions, with three or more field components,
can be characterized by $\mathbb{C}P^{N-1}$  topological invariants.
These are much less studied in condensed matter systems,
where high broken symmetries have been relatively rare.
Recently, the interest in these objects started to increase \cite{Garaud2011topological,Garaud_CP2_2013,kovrizhin2013multicomponent,garaud2014domain,lian2017spin,akagi2021isolated,zhang2022cp,akagi2021fractional,Amari2022CP2},  revealing very interesting properties. 

In our work we focus on superconducting systems.
Previously, it was shown that \skyr skyrmions are possible in a phenomenological Ginzburg-Landau model describing a three-band $s+\ii s$ superconductor \cite{Garaud2011topological,Garaud_CP2_2013}.
A three-band $s+\ii s$ superconductor \cite{Stanev2010,Carlstrom2011_lengthscales,Maiti2013,Boeker2017} can be described by a
three-component Ginzburg-Landau theory where
$U(1)\times U(1) \times U(1)$ symmetry is explicitly broken to $U(1)\times Z_2$ by intercomponent coupling \cite{Carlstrom2011_lengthscales,garaud2017microscopically}.
The Ginzburg-Landau-based studies in \cite{Garaud2011topological,Garaud_CP2_2013} suggest that \skyr skyrmions can form in $s+\ii s$ superconductors as meta-stable states, with a slightly higher energy per flux quanta than ordinary vortices. They could be excited by perturbations and be protected against decay by a potential energy barrier.
The experimental discovery of $s+\ii s$  superconductivity in Ba$_{\rm 1-x}$K$_{\rm x}$Fe$_2$As$_2$ was recently
reported \cite{Grinenko2017superconductivity,Grinenko2018s+is,Grinenko2021_state}. The evidence is based on the detection and analysis of spontaneous magnetic fields
from muon spin relaxation experiments \cite{Grinenko2018s+is,Vadimov2018}, spontaneous Nernst effect \cite{Grinenko2021_state}, and the existence of two phase transitions indicating two broken symmetries \cite{Bojesen2013time,Bojesen2014_phase,Grinenko2021_state}.
This is a strong motivation to investigate the existence of \skyr skyrmions in a fully microscopic model, not limited to temperatures near criticality. 

In this paper we report the existence of \skyr skyrmions in fully self-consistent solutions to a three-band Bogoliubov-de Gennes (BdG) model.
The model retains all microscopic degrees of freedom.
Moreover, the microscopic calculations allows us to determine the skyrmions' signatures in the local density of states.
\section{Model}

\begin{figure}[t]
    \centering
    \includegraphics[width=0.20\textwidth]{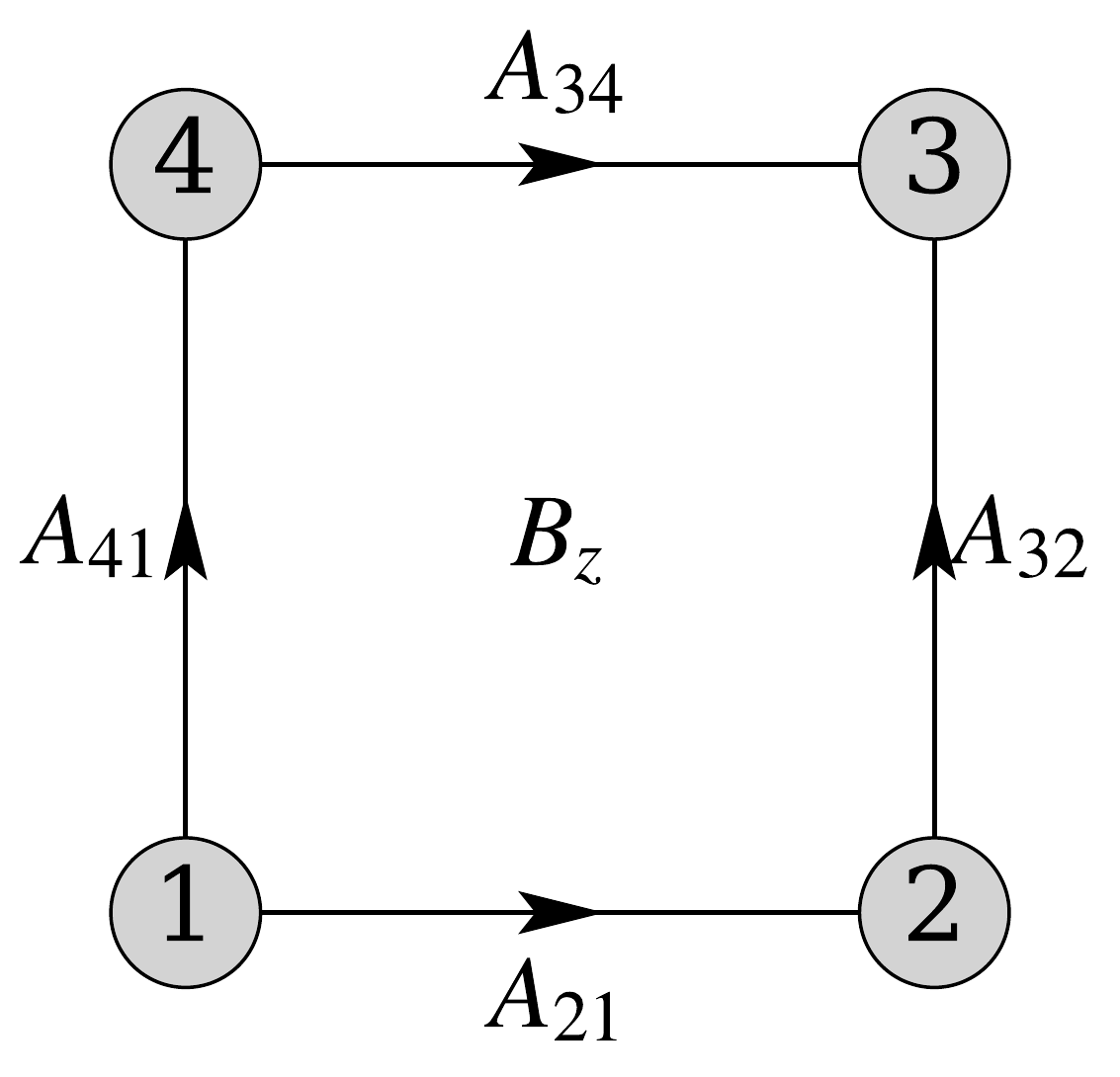}
    \caption{Illustration of the lattice description of the magnetic field (defined on the plaquette), and the vector potential (defined on the links).}\label{fig: latticeSetup}
\end{figure}

We begin by considering the three-component Hubbard model, defined on a two-dimensional square lattice, described by the microscopic Hamiltonian %
\begin{equation} \label{eq: full Hubbard hamiltonian}
\begin{aligned}
        H = & - \sum_{\alpha\sigma}\sum_{<ij>}\exp{\ii q A_{ij} }c^\dagger_{i\sigma\alpha}c_{j\sigma\alpha} \\ & -\sum_{i\alpha\beta}V_{\alpha\beta} c_{i\uparrow\alpha}^\dagger c_{i\downarrow\alpha}^\dagger c_{i\downarrow\beta} c_{i\uparrow\beta}.
\end{aligned}
\end{equation}
Here $<ij>$ denotes nearest neighbor pairs, and $c_{i \sigma \alpha}$ is the fermionic annihilation operator at position $i$, with spin $\sigma$ ($\sigma \in \lbrace \uparrow, \downarrow \rbrace$) and in band $\alpha$ ($\alpha \in \lbrace 1,2,3 \rbrace$). We are using a rescaled unit system, where the planar spatial coordinates are measured in units of the lattice spacing, and all energies are measured in units of the hopping parameter (for details see \appendixref{appendix: rescaling}).
The quartic interaction term, defined by $V_{\alpha \beta} = V_{\beta \alpha}^*$, allows Cooper pairs to form and tunnel between bands, yielding multiband superconductivity.
By performing the mean field approximation in the Cooper channel (for details see \appendixref{appendix: modelDerivation}), we obtain the mean field Hamiltonian
\begin{equation}\label{eq: mfHamiltonian}
\begin{aligned}
    \mathcal{H}= & -  \sum_{\sigma\alpha}\sum_{<ij>} \exp{\ii q A_{ij} } c^\dagger_{i \sigma \alpha } c_{j\sigma } \\
    & + \sum_{i \alpha} \left( \Delta_{i \alpha} c^\dagger_{\uparrow i \alpha} c^\dagger_{\downarrow i \alpha} +  \hc \right),
\end{aligned}
\end{equation}
where $\hc$ denotes Hermitian conjugation.
The interaction between the bands is embedded in the self-consistency equations for the gaps, which read
\begin{equation}\label{eq: selfConsistency}
    \Delta_{i \alpha} = \sum_\beta V_{\alpha \beta} \expval{c_{\uparrow i \beta} c_{\downarrow i \beta}}.
\end{equation}
The phase factor $\exp{\ii q A_{ij}}$ accounts for interaction with the magnetic vector potential $\Vec{A}$ through Peierls substitution \cite{Peierl,feynman2011mainly}, where
\begin{equation}
    A_{ij} = \int_j^i \Vec{A} \cdot \dd{\Vec{\ell}}.
\end{equation}
The magnetic field is also solved for self-consistently. The current from $j$ to $i$ generated by the fermions equals
\begin{equation}\label{eq: currentDensity}
\begin{aligned}
    J_{ij} & = - \expval{ \pdv{\mathcal{H}}{A_{ij}}} = \\
    & = -2q \sum_{\alpha\sigma} \Im{ \expval{c_{i\sigma\alpha}^\dagger c_{j \sigma \alpha}} \exp{\ii q A_{ij} }}.
\end{aligned}
\end{equation}
The current is defined on the links connecting nearest neighboring sites, and we discretize the vector potential in the same way, as shown in \figref{fig: latticeSetup}. This results in having the magnetic field $B_z$ defined on the lattice plaquettes, and equal to the discrete curl of the vector potential. Following \figref{fig: latticeSetup}, we have $B_z  = A_{21} + A_{32} - A_{34} - A_{41}$.
Following this convention, we can write the magnetic field energy as 
\begin{equation}
    F_{\textrm{mag}} = \frac{1}{2}\sum_{\rm{plaquettes}}  B_z^2,
\end{equation}
where the sum is carried out over all the plaquettes. 
The discrete version of Maxwell equation $\curl\Vec{B} -\Vec{J}=\Vec{0}$ is
\begin{equation}\label{eq: Aeom}
    \pdv{F_{\magnetic}}{A_{ij}} + \expval{ \pdv{\mathcal{H}}{A_{ij}}} =0.
\end{equation}
The free energy associated with the tight binding Hamiltonian, up to a constant, equals
\begin{equation}\label{eq: FH}
    F_H = \sum_{i}\Vec{\Delta}_i^\dagger V^{-1}\Vec{\Delta}_i -k_\B T\Tr\ln\qty(\ee^{-\beta \mathcal{H}}+1),
\end{equation}
such that the total free energy is $F = F_\magnetic + F_H$.

 The self-consistency equations in \eqref{eq: selfConsistency}, along with the Maxwell equation for the magnetic field in \eqref{eq: Aeom}, are solved numerically by employing an iterative scheme based on the Chebyshev spectral expansion method \cite{covaci2010efficient,nagai2012efficient,weisse2006kernel}. For more details on the iterative scheme, see \appendixref{appendix: numericalMethods}.

\subsection{Gauge and time-reversal symmetries}

Having introduced coupling between the fermions and the magnetic vector potential using Peierls substitution, the system is now invariant under the gauge transformation
\begin{equation}
    \begin{aligned}
        c_{i\sigma\alpha} & \mapsto \exp{\ii \chi_i} c_{i\sigma\alpha}, \\
        A_{ij} & \mapsto A_{ij} + \frac{\chi_i - \chi_j}{q}, \\
        \Delta_{i\alpha} & \mapsto \exp{2 \ii  \chi_i} \Delta_{i\alpha}.
    \end{aligned}
\end{equation}
This transformations does not alter the magnetic field or the phase differences between the superconducting gaps. In this notation, the magnetic flux quantum, associated with one vortex, is equal to $\Phi_0 = \pi/q$.

Suppose that $(\Vec{\Delta}, \Vec{A})$ is a self-consistent solution to our equations. By symmetry, $(\Vec{\Delta}^*, -\Vec{A})$ is also a solution.
The $s+\ii s$ state in three-band system corresponds to the case where the phase differences between the individual gaps are different from zero and $\pi$. This state spontaneously breaks time-reversal symmetry, since $\Vec{\Delta}$ and $\Vec{\Delta}^*$ are distinct states (corresponding to $s+\ii s$ and $s-\ii s$ superconducting states respectively). These two time-reversed states cannot be transformed into each other by a gauge transformation.
\subsection{Skyrmionic index} \label{sec: skyrmionic index}

In this section, we introduce the classification of topological excitations in terms of their topological index. The model contains three complex valued fields $\Vec{\Delta} = (\Delta_1, \Delta_2, \Delta_3)^\T$, defined on a square lattice. We can write $\Vec{\Delta} = \rho \Vec{Z}$, where $\Vec{Z} = (Z_1, Z_2, Z_3)^\T$ satisfies $|\Vec{Z}|^2=1$ and contains information about the relative phases and magnitudes of the three superconducting gaps. The total density $\rho$ and the relative phases and magnitudes contained in $\Vec{Z}$ are gauge invariant. If we assume that $\rho > 0$ (that is, the superconducting gaps never vanish simultaneously), the space associated with the relative phases and magnitudes is the complex projective plane $\mathbb{C}P^2$. 

For a lattice model, the topological index $\mathcal{Q} \in \mathbb{Z}$, associated with the complex projective space $\mathbb{C}P^{2}$, has been derived in \cite{Berg_topological_number_lattice}, and is defined as a sum over the plaquettes, where each plaquette contains two signed triangles. Similarly as in \figref{fig: latticeSetup}, let 1, 2, 3 and 4 denote the vertices associated with a plaquette, in a counter-clockwise order. One possible choice for the two signed triangles is 123 and 134.
The skyrmionic charge density  $\rho_{\mathbb{C}P^{2}}({\rm{ABC}})$ associated with a signed triangle ABC satisfies
\begin{equation}
    \ee^{\ii 2\pi \rho_{\mathbb{C}P^{2}}({\rm{ABC}})} = \frac{\Tr{P_{\rm{A}}P_{\rm{B}}P_{\rm{C}}}}{\sqrt{ \Tr{P_{\rm{A}}P_{\rm{B}}} \Tr{P_{\rm{B}}P_{\rm{C}}} \Tr{P_{\rm{C}}P_{\rm{A}}} }},
\end{equation}
where the matrices $P$, defined as $P_{\alpha \beta} = Z_\alpha Z_\beta^*$, contain the relative phase and magnitudes of the complex fields at the different vertices \footnote{Note that here we are using a different sign convention than in \cite{Berg_topological_number_lattice}, such that $\mathcal{Q}>0$ for topological excitations associated with positive magnetic flux (for positive charge $q$ and for this order of $s\pm \ii s$ domains)}. Explicitly, the topological index reads
\begin{equation}
    \mathcal{Q} = \sum_{\textrm{plaquettes}} \left[ \rho_{\mathbb{C}P^{2}}(123) + \rho_{\mathbb{C}P^{2}}(134) \right].
\end{equation}
The corresponding topological index for a continuum Ginzburg-Landau model for a three-component superconductor was discussed in \cite{Garaud_CP2_2013}.

\section{Ground state and critical temperature} \label{sec: ground state}

The ground state is found by considering the case when the superconducting gaps $\Vec{\Delta} = \qty(\Delta_1, \Delta_2, \Delta_3)^\T$ are uniform, in the absence of any magnetic field. The self-consistency equations in the multi-component case are a straightforward extension of the single component case, and read 
\begin{equation} \label{eq: self-consistency equation uniform}
    \Delta_\alpha = \sum_\beta V_{\alpha \beta} I(|\Delta_\beta|^2) \Delta_\beta,
\end{equation}
where
\begin{equation} \label{eq: definition of integral uniform}
    I(|\Delta|^2) = \int_\BZ \frac{\dd{k_x}\dd{k_y}}{(2\pi)^2} \frac{ \tanh \left\lbrace \frac{1}{2 k_\B T} E (\Vec{k}, |\Delta|^2) \right\rbrace }{2 E(\Vec{k}, |\Delta|^2) },
\end{equation}
and $E (\Vec{k}, |\Delta|^2) = \sqrt{\xi(\Vec{k})^2 + |\Delta|^2 }$ is the dispersion relation in the presence of the superconducting gap $\Delta$, with $\xi (\Vec{k}) = -2\cos k_x -2\cos k_y$ being the dispersion relation in the absence of superconductivity. The integral is carried out over the first Brillouin zone $[0, 2\pi) \times [0, 2\pi)$. In order to find the critical temperature, at which the superconducting gap vanishes continuously, we can expand \eqref{eq: self-consistency equation uniform} to the lowest order in the gap magnitude, resulting in the linearized gap equation
\begin{equation} \label{eq: uniform linearised gap equation}
    \Vec{\Delta} = I(0) V \Vec{\Delta}.
\end{equation}
It is clear that the critical temperature is determined by the largest eigenvalue $V_{\max}$ of the pairing potential matrix $V$, such that $1=I(0)V_{\max}$. The relative phases and magnitudes of the components $\Delta_\alpha$ at criticality can be inferred from the corresponding eigenvector.

\begin{figure}
    \centering
    \includegraphics[width=1.0\columnwidth]{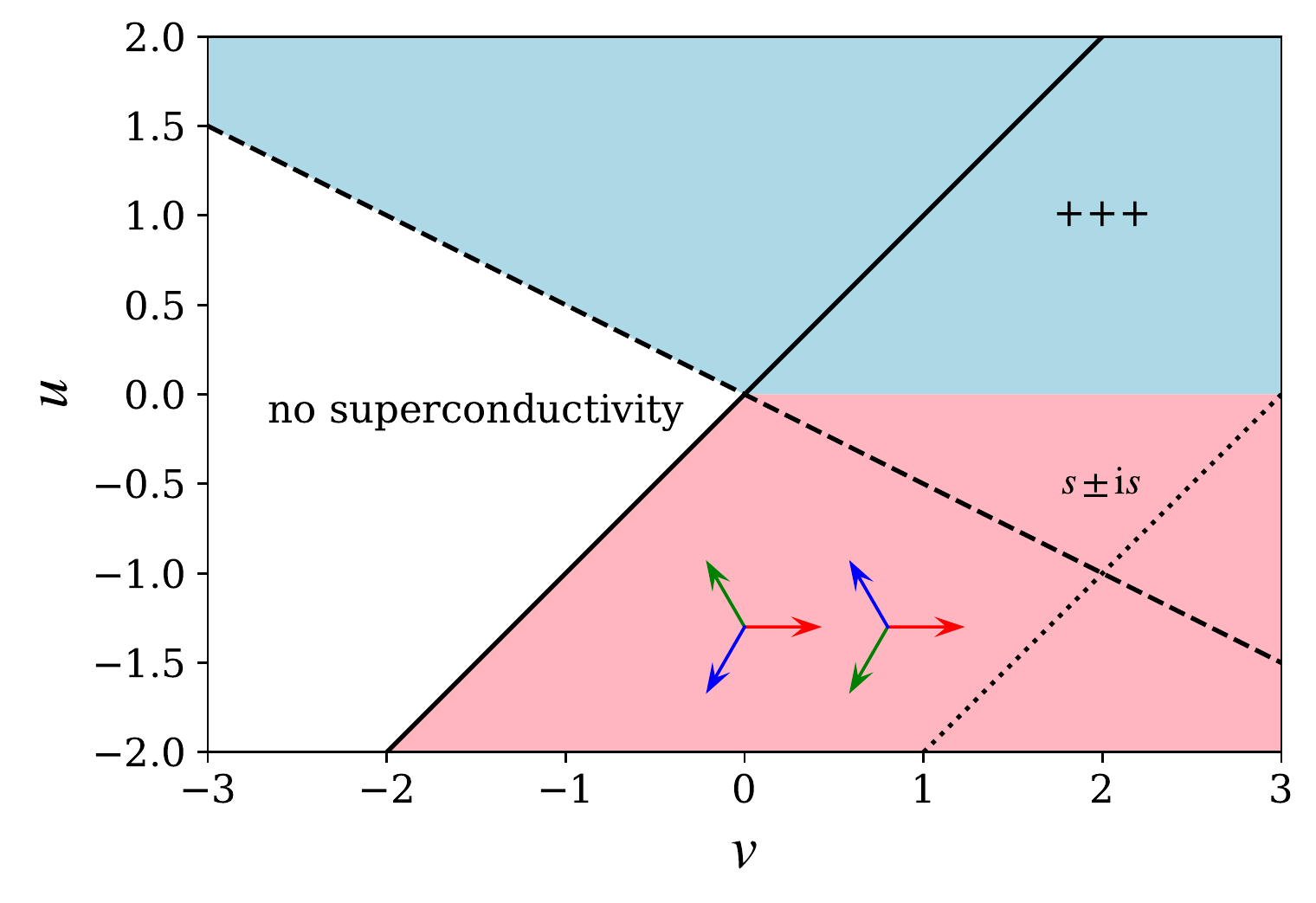}
    \caption{Phase diagram for the superconducting ground state, where $v$ is the pairing potential within bands, and $u$ is the interaction pairing between different bands, as in \eqref{eq: pairing potential matrix}. When $u$ is positive, the optimal phase difference between the bands is $0$, while for negative $u$, the optimal phase difference is $\pi$, resulting in frustration and spontaneous time reversal symmetry breaking (illustrated by the two arrow configurations). The full (dashed) line indicates when the pairing potential matrix is singular, i.e. when the eigenvalue $V_-$ ($V_+$) is zero. No superconductivity is present when both eigenvalues are negative. 
    The dotted line marks where we perform the investigation of skyrmions.}
    \label{fig: ground state phase diagram}
\end{figure}

We now consider the pairing potential matrix $V_{\alpha \beta}$, written in matrix form as
\begin{equation} \label{eq: pairing potential matrix}
    V = 
    \begin{pmatrix}
        v & u & u \\
        u & v & u \\
        u & u & v
    \end{pmatrix},
\end{equation}
where $v$ is the pairing potential within each band, also called \textit{intraband} pairing, while $u$ is the interaction pairing between different bands, i.e., \textit{interband} interaction. Diagonalizing the pairing potential matrix $V$ in \eqref{eq: pairing potential matrix}, gives us two eigenvalues, $V_+ = v+2u$ and $V_-=v-u$. The eigenvector associated with $V_+$ is $\Vec{v}_{+++} = (1,1,1)^\T /\sqrt{3}$, corresponding to zero phase difference between all three components. The eigenvalue $V_-$ is degenerate, and a possible choice for the two orthonormal associated eigenvectors is
\begin{equation}
    \Vec{v}_{s + \ii s} = \frac{1}{\sqrt{3}}
    \begin{pmatrix}
        1 \\ \omega_3 \\ \conj{\omega}_3
    \end{pmatrix},
    \quad
    \Vec{v}_{s - \ii s} = \frac{1}{\sqrt{3}}
    \begin{pmatrix}
        1 \\ \conj{\omega}_3 \\ \omega_3
    \end{pmatrix},
\end{equation}
where $\omega_3 = \ee^{ 2\pi \ii / 3}$ is the third root of unity. The eigenvectors $\Vec{v}_{s\pm\ii s }$ have non-zero phase differences between all three components, and are each others complex conjugates. These three eigenvectors form a complete orthonormal basis, and consequently the gap vector $\Vec{\Delta}$ can be expressed as
\begin{equation}
    \Vec{\Delta} = \Delta_{s + \ii s}\Vec{v}_{s + \ii s} + \Delta_{s - \ii s}\Vec{v}_{s - \ii s} + \Delta_{+++} \Vec{v}_{+++}.
\end{equation}
We are interested in the case when $u$ is negative, such that opposite signs (phase difference $\pi$) of the superconducting gaps is favored, and $V_-$ is the largest eigenvalue. 
In the case of three components, this results in phase frustration, since the optimal phase difference $\pi$ cannot be satisfied for all pairs. The resulting ground state $\Vec{\Delta} \propto \Vec{v}_{s\pm \ii s}$ spontaneously breaks time reversal symmetry. It is important to note that it is not possible to infer directly from the linearized gap equation the ground state for temperatures below the critical temperature. For example, the state $\Vec{v}_{++-} = (1,1,-2)^\T / \sqrt{5}$, is also an eigenvector with eigenvalue $V_-$. However, studying the associated free energies below the critical temperature shows that the $s\pm \ii s$ state is energetically more favorable. The phase diagram is shown in \figref{fig: ground state phase diagram}.

At the critical lines $V_\pm = 0$, the pairing potential matrix is singular, and the free energy in \eqref{eq: FH} seems ill-defined. However, this is not the case. For example, when $V_+=0$, the projection of $\Vec{\Delta}$ onto the corresponding eigenspace is exactly zero, and the three superconducting gaps $\Vec{\Delta}$ can always be written as a linear combination of only two order parameters: $\Delta_{s+\ii s}$ and $\Delta_{s-\ii s}$. Similarly, when $V_-=0$, the full three-band system is described by a single order parameter $\Delta_{+++}$. It should be noted that this is not only for the ground state, but holds locally for every inhomogeneous gap configuration.

\section{Skyrmion solutions} \label{sec: investigation of skyrmions}

Similarly as for conventional superconductors, the three-band model in this paper, described by three complex fields $\Vec{\Delta} = (\Delta_1, \Delta_2, \Delta_3)^\T$ can host topological excitations, in the form of vortices. The most simple vortex solution would be the composite vortex, where all three complex fields vanish simultaneously at a joint vortex core. Overlapping vortex cores are favored by the electromagnetic and the Josephson coupling  \cite{Babaev2002_vortices}. The topological invariant associated with this solution is the winding number of the complex fields. 
Fractional vortices, i.e., configurations with winding in only one of the three complex fields have been found stable \cite{Babaev2002_vortices,garaud2014domain}, but energetically unfavorable compared to composite vortices.
The skyrmion solutions, shown previously using Ginzburg-Landau theory for an $s+\ii s$ superconductor \cite{Garaud_CP2_2013,Garaud2011topological}, are situations in which the vortex cores can fractionalize, i.e., the vortex cores of the three complex fields do not overlap. The fractionalized vortices populate domain walls.
At the domain wall the interaction between fractional vortices is repulsive, since phase differences between components have energetically unfavorable values. Hence, the domain wall is stabilized against  collapse due to the Josephson interaction together with the magnetic repulsion between the fractional vortices.
In these cases, the solution is associated with a topological skyrmionic index $\mathcal{Q}$, as described in \secref{sec: skyrmionic index}. In this section, we  search for such solutions and investigate their properties in a fully microscopic model.

A full phase diagram exploration of these properties is unfeasible. However, by setting $V_- = v-u = 3$, while varying both $u$ and $v$, we can explore regimes where the interaction between the bands is weak ($|u|$ is small) and where the interaction between the bands is dominant ($|u|$ is large). Since $V_-$ is constant along this line, so is the critical temperature ($k_\B T_c \approx 0.46$). This line is marked as the dotted line in \figref{fig: ground state phase diagram}. We perform simulations at $T=T_c / 2$, where Ginzburg-Landau theory is unreliable.
The numerical solutions we present are obtained from simulating finite-sized systems that are large, such that the skyrmions are sufficiently far away from the boundaries in order to discard any mesoscopic effects.

\subsection{Skyrmion structure: $\mathcal{Q}=5$ example} \label{sec: first example Q=5}
\begin{figure}[t]
    \centering
    \includegraphics[width=1.0\columnwidth]{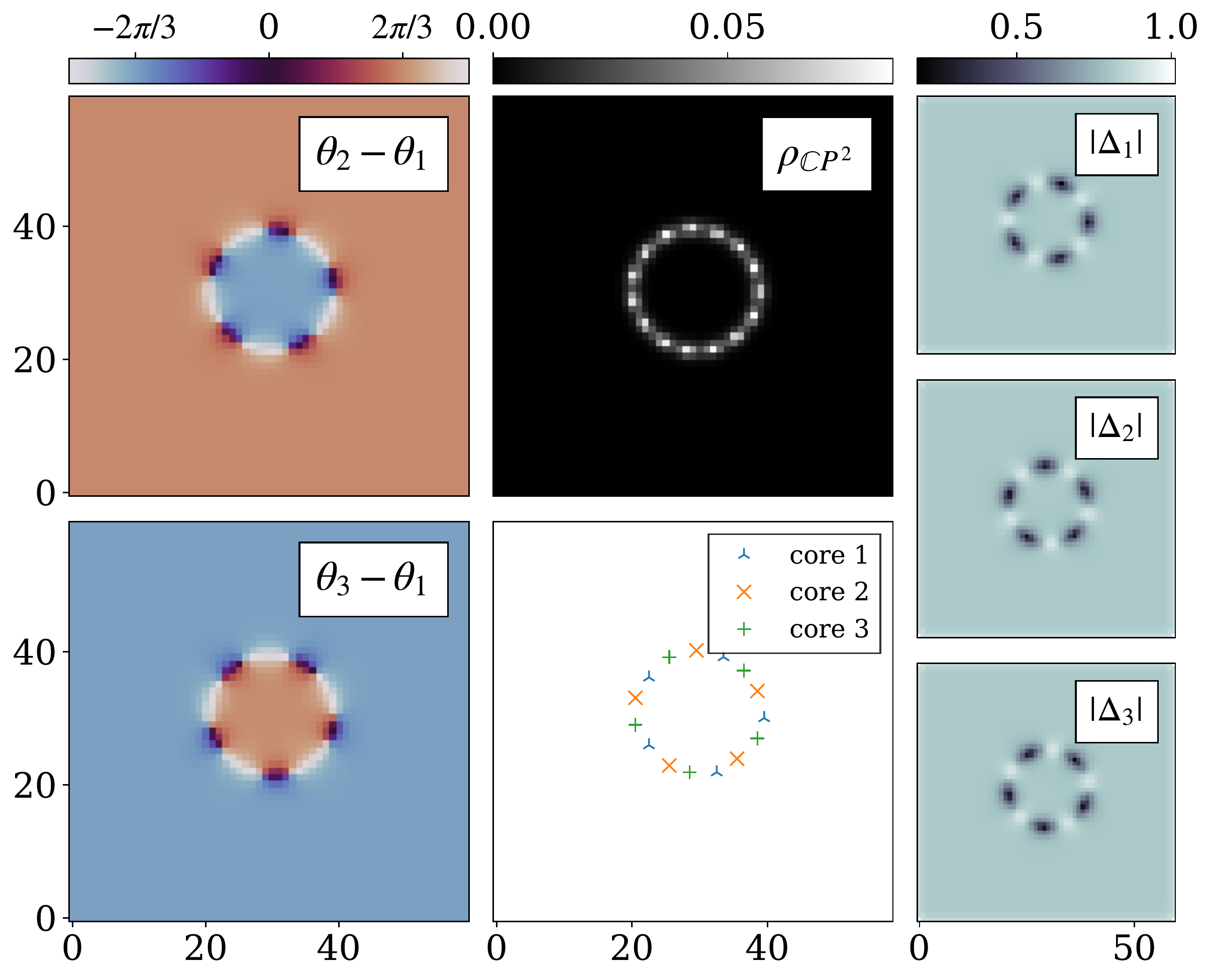}
    \caption{Example of skyrmion with $\mathcal{Q}=5$. The left column shows the two phase differences $\theta_2-\theta_1$ and $\theta_3-\theta_1$, showing that the skyrmion forms a circular domain wall, stabilized by the vortices. The second column shows the skyrmion charge density $\rho_{\mathbb{C}P^{2}}$ and the fractionalization of the vortex cores. The third column shows the absolute value of each superconducting gap. We use intraband pairing $v=2.5$, interband pairing $u=-0.5$ and effective charge $q=0.15$.}
    \label{fig: example Q=5 skyrmion}
\end{figure}

Let us begin by studying the structure of a single skyrmion with skyrmionic topological charge $\mathcal{Q}=5$, shown in \figref{fig: example Q=5 skyrmion}. We set the effective charge $q=0.15$, the pairing matrix diagonal $v=2.5$ and off-diagonal $u=0.5$. The skyrmion is an excitation on top of the $s + \ii s$ ground state, where the phase differences are $\theta_2-\theta_1 = 2\pi/3$ and $\theta_3-\theta_1 = -2\pi/3$. The skyrmion consists of 5 fractional vortices in each superconducting band, that is with non-overlapping vortex cores, forming a closed ring. Inside the ring, the phase differences change sign and become an $s- \ii s$ superconducting state. 
Note also that the vortices form a repeating pattern along the domain wall. The cores are ordered in a $123123\ldots$ fashion (counter-clockwise). The cyclic ordering $132132\ldots$ is not stable, unless the vortices are replaced with anti-vortices, or if the $s\pm \ii s$ domains are interchanged. 
A preferential sequence for ordering of fractional vortices, i.e. the skyrmion's chirality, was also found in Ginzburg-Landau models \cite{Garaud_CP2_2013}.
 
\begin{figure}[t]
    \centering
    \includegraphics[width=1.0\columnwidth]{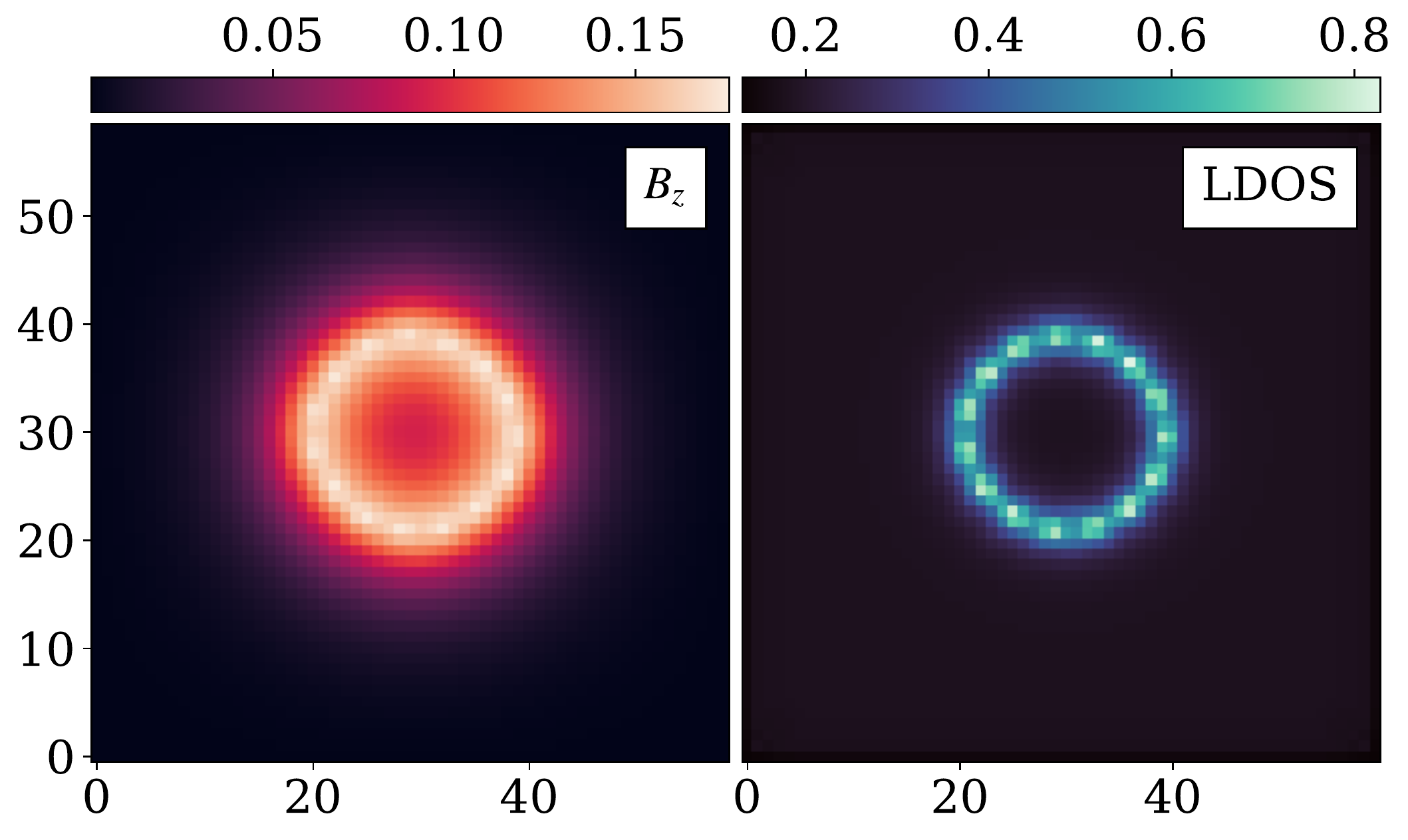}
    \caption{
    Magnetic field $B_z$ and local density of states (LDOS) at Fermi energy for the $\mathcal{Q}=5$ skyrmion in \figref{fig: example Q=5 skyrmion}. Both quantities are localized at the domain wall ring and the fractional vortex cores.
    }
    \label{fig: example Q=5 skyrmion B and LDOS}
\end{figure}

Our microscopic model allows us to calculate the local density of states (LDOS) at the Fermi energy. It  is shown in \figref{fig: example Q=5 skyrmion B and LDOS} together with the magnetic field generated by the skyrmion. The LDOS at energy $\epsilon$ is calculated using the quasi-particle wave-functions $u_{\uparrow n i}$ and $v_{\downarrow n i}$ with eigenenergy $E_n$ as
\begin{equation} \label{eq: LDOS formula}
    {\rm{LDOS}}_i(\epsilon) = -\sum_n \Big( |u_{\uparrow n i}|^2f'(E_n-\epsilon) + |v_{\downarrow n i}|^2f'(E_n+\epsilon) \Big),
\end{equation}
where $f(x) = (1+\ee^{\beta x})^{-1}$ is the Fermi-Dirac distribution and $\beta = 1/(k_\B T)$. In this model, the Fermi energy is set at $\epsilon=0$. Both the magnetic field and the LDOS are localized at the fractional vortex cores.
The possibility of resolving the magnetic field peaks of individual vortex cores depends on the skyrmion size in relation to the magnetic field length scale.
In the following sections, we investigate how the skyrmion size and excitation energy depend on the interband pairing strength, magnetic lengthscale and skyrmion index.

\subsection{The influence of interband pairing strength} \label{sec: interband scaling}

\begin{figure}[t]
    \centering
    \includegraphics[width=1.0\columnwidth]{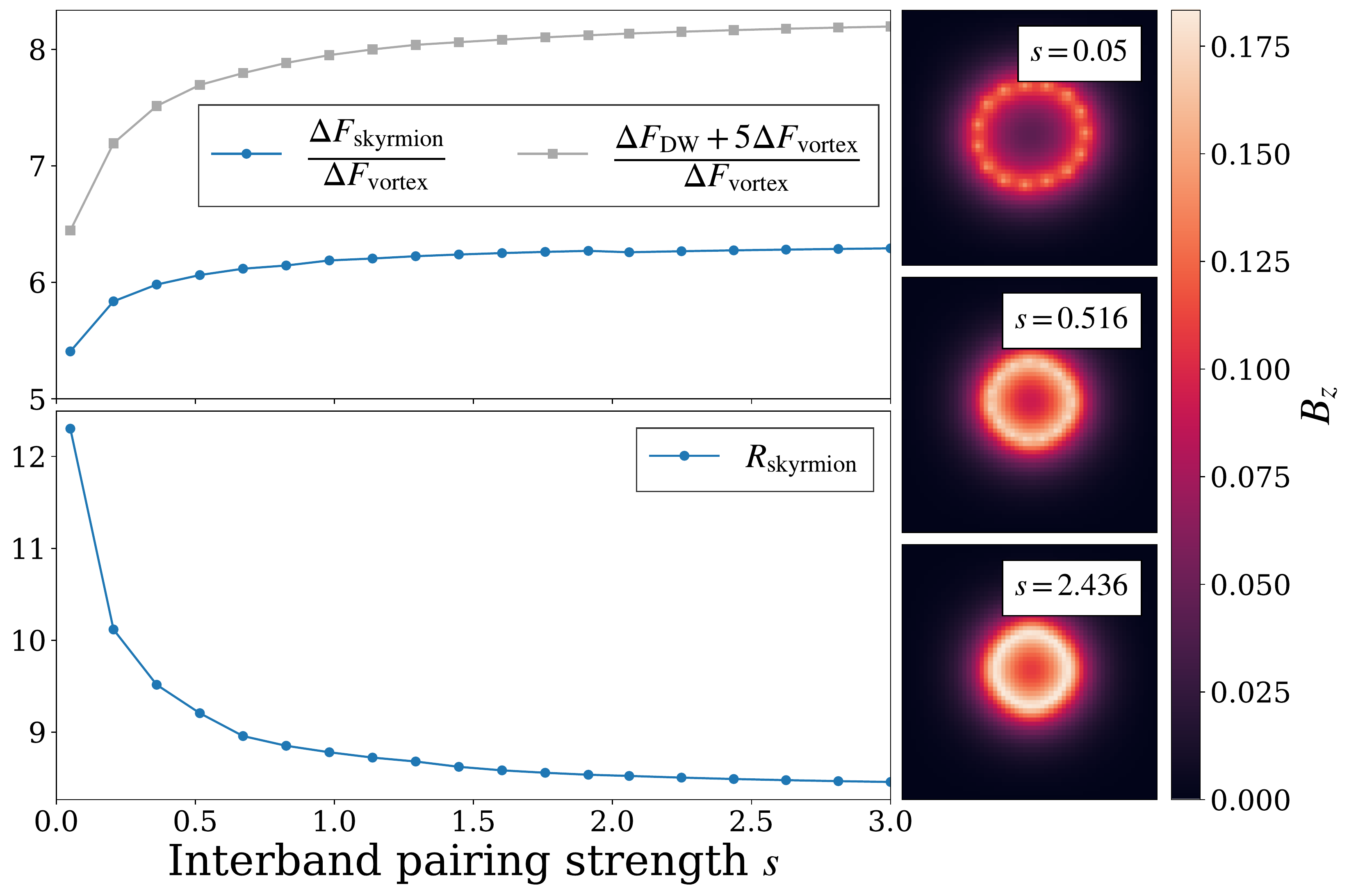}
    \caption{Excitation energy and skyrmion radius for a $\mathcal{Q}=5$ skyrmion, for different strengths of the interband pairing, parametrized by $s$. Here $s=0$ corresponds to no interaction between bands, and $s\to \infty$ corresponds to interband-dominated pairing. The excitation energy is given in units of the vortex excitation energy. 
    The gray curve shows the energy associated with 5 vortices separated from a ring domain wall with  the radius of the skyrmion. The fact that the excitation energy of the skyrmion is smaller than the sum of the excitation energy of its constituents, shows that domain walls and vortices bind together.
    The right column shows the magnetic field for some specific values of $s$. We use effective charge $q=0.15$, intraband pairing $v=3-s$ and interband pairing $u=-s$.} \label{fig: energy and radius vs s}
\end{figure}

Having demonstrated an example of a stable skyrmion, we now explore its properties as we change the interaction between the bands. To be specific, we study how the skyrmion excitation energy and size change. As previously mentioned, we fix $v-u=3$ and change both the interband $u$ and the intraband $v$ pairing simultaneously. We parametrize this change as $u=-s$ and $v=3-s$ and consider positive values of $s$. That is, $s=0$ corresponds to no interaction between the bands, while large $s$ corresponds to interband-dominated pairing. We compute the area of the skyrmion as
\begin{equation}
    A = \frac{1}{2}\sum_i \left( 1-\frac{2}{\sqrt{3}} \sin (\theta_2-\theta_1)_i \right),
\end{equation}
and the associated radius $R_\skyrmion = \sqrt{A/\pi}$. 
To compute the skyrmion excitation energy $\Delta F_\skyrmion$, we calculate the total free energy of the system, form which we subtract the free energy of the ground state. 
By doing so, we are able to cancel the free energy contributions coming from the boundary and uniform zones of the gap. 
The excitation energy and skyrmion radius for a $\mathcal{Q}=5$ skyrmion are shown in \figref{fig: energy and radius vs s}. When the interband pairing is weak, the excitation energy is small, and the size of the skyrmion grows. When the interband pairing is strong, the skyrmion shrinks and the excitation energy increases, both approaching constants in the limit of interband-dominated pairing. The excitation energy is given in units of the excitation energy of a composite vortex. We see that in the limit of weak interband pairing, that is $s \to 0$, the skyrmion excitation energy approaches the excitation energy associated with $\mathcal{Q}$ vortices. In this limit of weak interband interaction, the skyrmion radius grows, allowing for the peaks of the magnetic field associated with each fractional vortex to be resolvable. In the limit of interband-dominated pairing, the skyrmion shrinks, such that the field from each fractional vortex no longer can be resolved. 
In \figref{fig: energy and radius vs s} we show, for comparison, the excitation energy associated with the constituents of the skyrmion: the domain wall and the 5 composite vortices. The excitation energy of the domain wall is calculated as the excitation energy per unit length of a straight domain wall, multiplied with the skyrmion circumference. The fact that the excitation energy of the skyrmion is smaller than the excitation energy of a domain wall and vortices separated, shows that skyrmions are formed from an attractive interaction between domain walls and vortices.  

\subsection{The influence of the magnetic lengthscale} \label{sec: magnetic length scaling}

\begin{figure}[t]
    \centering
    \includegraphics[width=1.0\columnwidth]{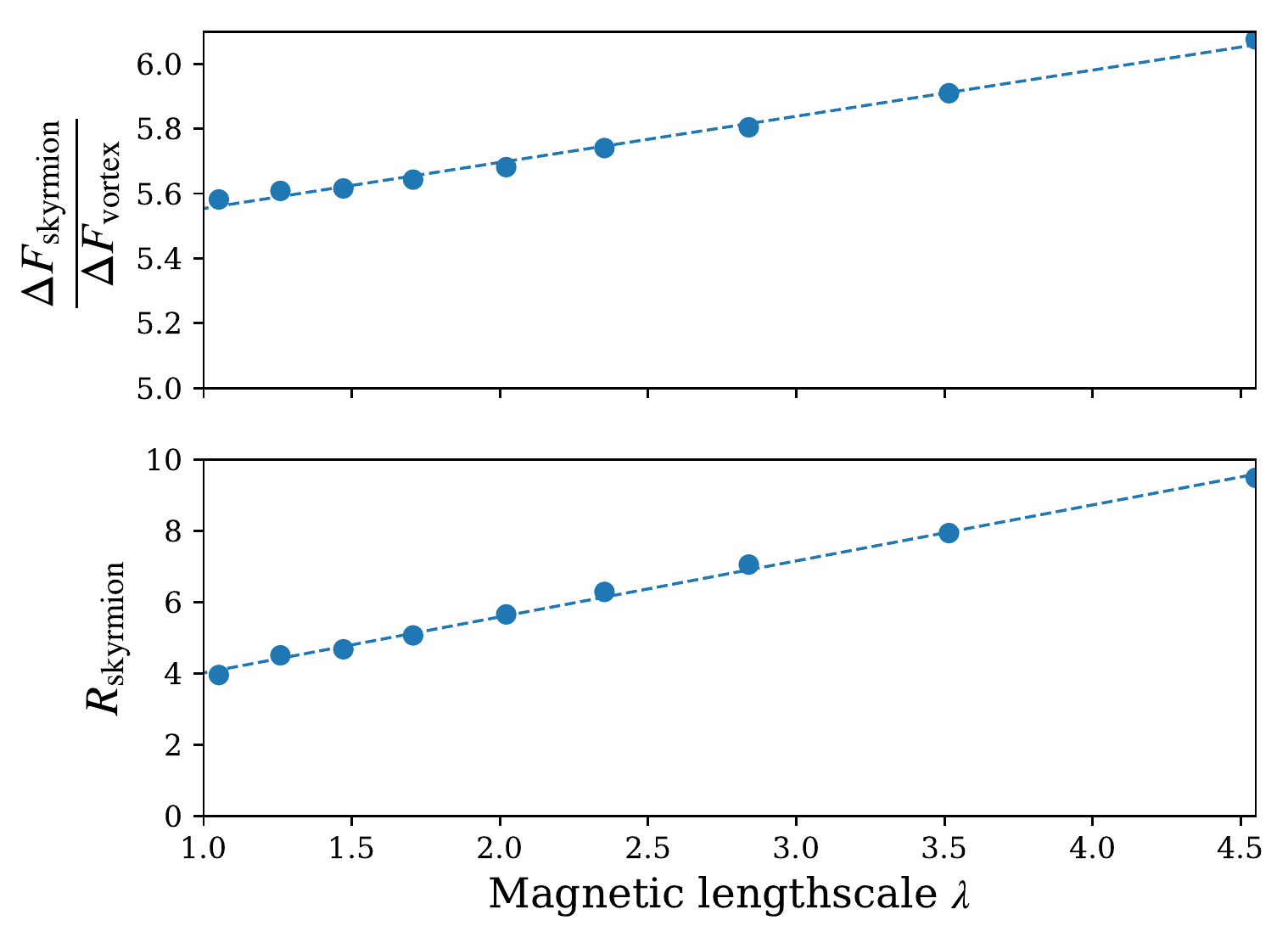}
    \caption{Excitation energy (in units of vortex excitation energy) and radius for a $\mathcal{Q}=5$ skyrmion, for various values of the magnetic lengthscale $\lambda$ associated with a composite vortex. In the considered parameters range, both quantities increase linearly with $\lambda$.  We use intraband pairing $v=2.5$ and interband pairing $u=-0.5$.} \label{fig: magnetic lengthscale comparison}
\end{figure}

In the previous examples, we have used a constant value of the effective charge $q$. In this section, we explore how changing $q$ affects the skyrmion size and energy. For vortices in one-band superconductors, the magnetic field decays on a lengthscale $\lambda$, which depends on the effective charge $q$. For example, in Ginzburg-Landau theory, $\lambda$ is proportional to the reciprocal of $q$.
In isotropic multiband models, the tail of the vortex magnetic field can be well approximated by $\ee^{-x/\lambda}$, where $\lambda$ is the lengthscale of interest. We consider values of the effective electric charge $q \in [0.144, 0.5]$, for which we compute the excitation energy and magnetic lengthscale for a single composite vortex, in addition to the excitation energy and radius of a $\mathcal{Q}=5$ skyrmion. The gathered data is presented in \figref{fig: magnetic lengthscale comparison}, where we show how the excitation energy and size of the skyrmion changes as the magnetic lengthscale is altered. 
Within the considered parameter regime, both quantities increase linearly with the magnetic lengthscale $\lambda$. Such a behavior is expected. For large $\lambda$, the repulsion between the fractional vortices is dominated by the magnetic repulsion. This explain why the skyrmion radius grows linearly with $\lambda$. The vortex excitation energy scales logarithmically with $\lambda$ (see e.g. \cite{svistunov2015superfluid}). 
The skyrmion excitation energy can be partitioned into three parts: the vortex energy, the domain wall energy and their interaction energy. The domain wall energy scales linearly with the skyrmion radius, which in turn scales linearly with $\lambda$. This domain wall-associated energy is dominant for larger $\lambda$ and explains the linear scaling of the excitation energy.
This means that when the magnetic penetration length is large, the excitation energy of flux-carrying skyrmions can be significantly larger than the excitation energy of the corresponding number of flux-carrying vortices. This disfavors the formation of skyrmions in materials with large magnetic penetration length.
Even though our data follows this expected behavior, it is unclear whether our range of magnetic lengthscales is sufficiently wide to extrapolate the trend for even larger values of $\lambda$. Numerical limitations on the system size do not allow us to reliably compute skyrmion properties in the BdG model for very large $\lambda$ (that is small $q$), and further investigations are necessary to confirm this scaling behavior in this limit.

\subsection{Energy and size scaling with $\mathcal{Q}$} \label{sec: Q scaling}

\begin{figure}[t]
    \centering
    \includegraphics[width=1.0\columnwidth]{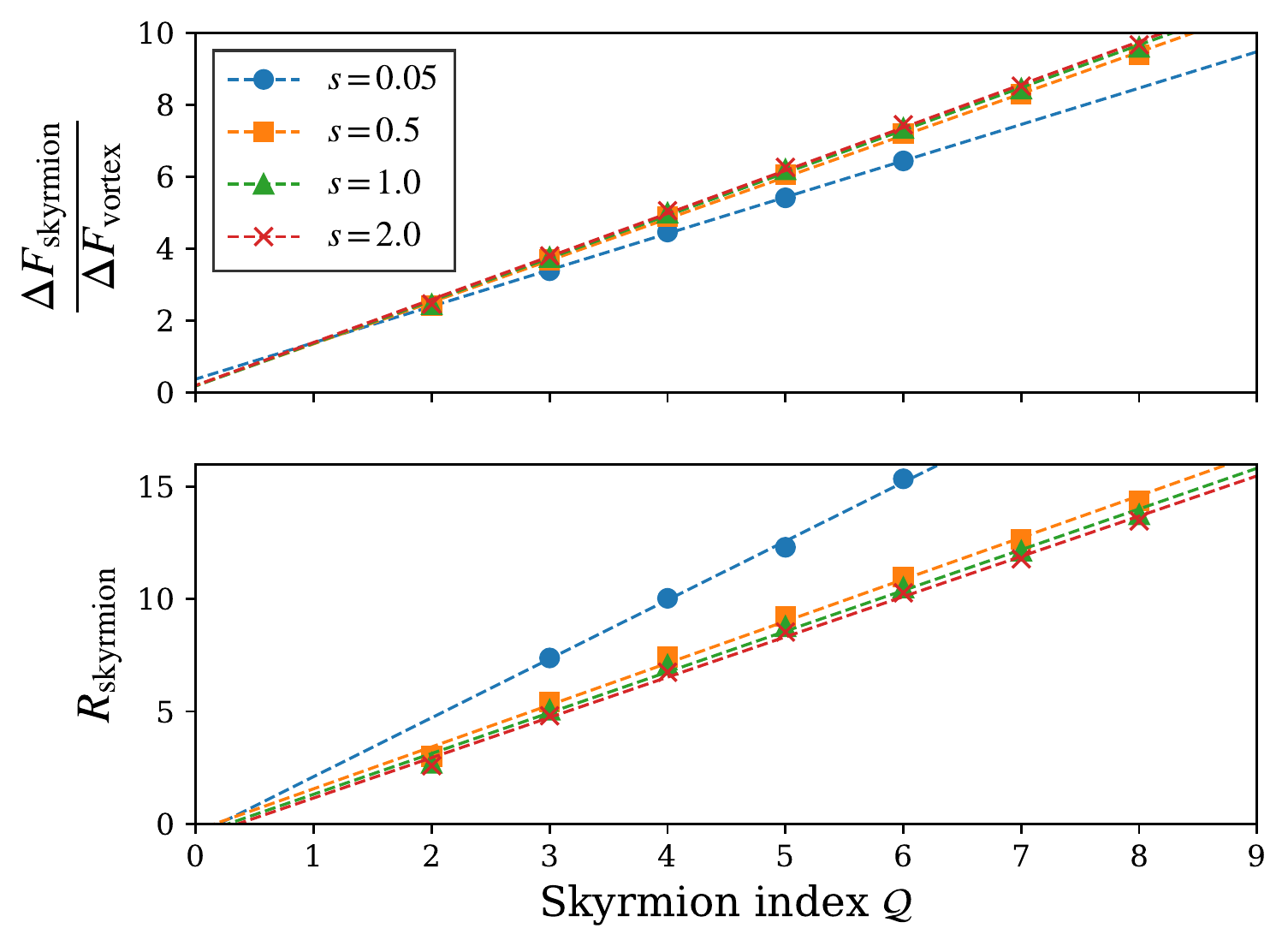}
    \caption{Skyrmion excitation energy (in units of vortex excitation energy) and skyrmion radius for increasing skyrmionic index $\mathcal{Q}$, for different interband pairing strengths $s$. Both quantities scale linearly with $\mathcal{Q}$. Increasing the interband pairing results in smaller skyrmions with larger excitation energies, as shown in \figref{fig: energy and radius vs s} for $\mathcal{Q}=5$ (we use the same parameters as in \figref{fig: energy and radius vs s}).} \label{fig: energy and radius vs Q}
\end{figure}

 Let us compute the skyrmion excitation energy and the skyrmion radius, for different skyrmionic indices $\mathcal{Q}$. The result is shown in \figref{fig: energy and radius vs Q}, where we consider different magnitudes of the interband pairing, controlled by the parameter $s$, which we introduced previously. We see that both the excitation energy and the radius grow linearly with $\mathcal{Q}$. Increasing the interband pairing strength (increasing $s$) results in smaller radii and larger excitation energies, as shown already for $\mathcal{Q}=5$ in \figref{fig: energy and radius vs s}. Note that for the weakest interaction considered, $s=0.05$, the stability of the skyrmion is achieved only for topological indices larger than $\mathcal{Q}=2$ .

\begin{figure*}
    \centering
    \includegraphics[width=1.0\textwidth]{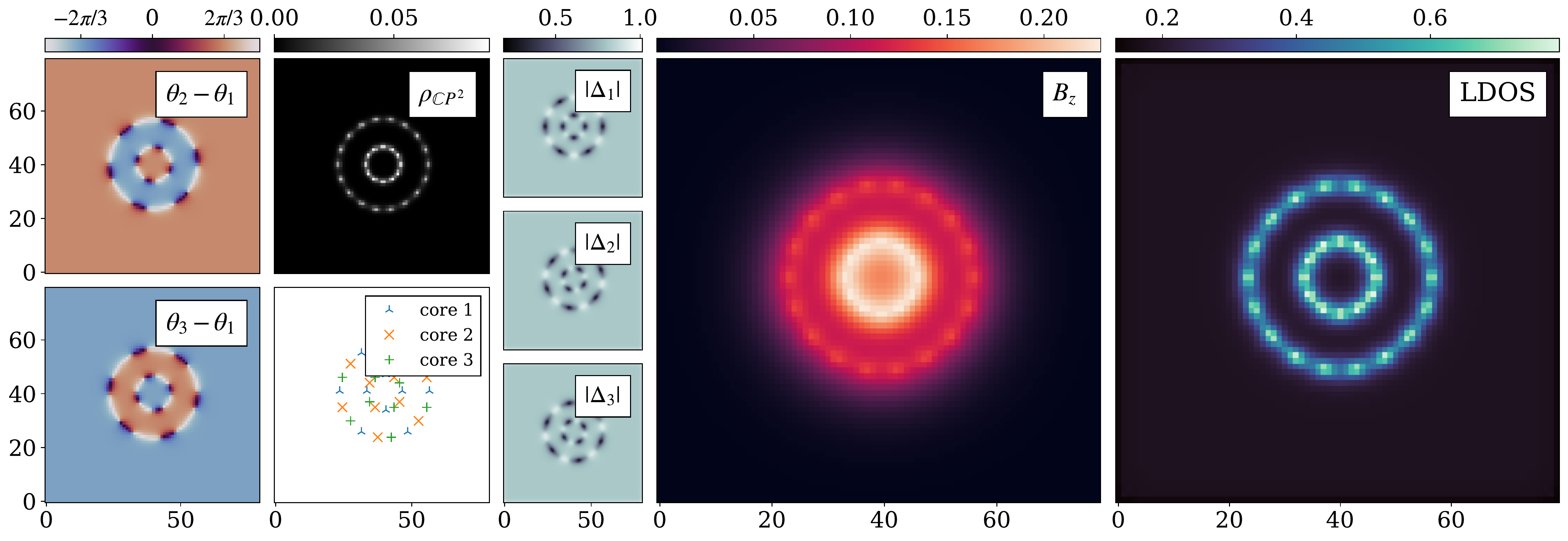}
    \caption{Example of a concentric skyrmion with $Q=10$, where the inner ring has 4 fractional vortices in each band, and the outer ring has 6. We use the same parameters and show the same quantities as in \figref{fig: example Q=5 skyrmion} and \figref{fig: example Q=5 skyrmion B and LDOS}.} \label{fig: concentric skyrmion}
\end{figure*}

\subsection{Concentric skyrmions} \label{sec: concentric}

In the previous examples, we considered different skyrmions with different topological indices $\mathcal{Q}$ by changing the number of vortices localized at the domain wall. However, it is also possible to change the number of domain walls. In this section, we show that these domain walls can even be concentric. 
Co-centered skyrmions have previously been reported using the Ginzburg-Landau model for $s+\ii s$ superconductors \cite{Garaud_CP2_2013}. Similar composite skyrmions have been shown to appear in chiral magnets \cite{Rybakov_Skyrmion_sacks}, and liquid crystals \cite{Foster2019}. In \figref{fig: concentric skyrmion} we show an example of a concentric skyrmion with $\mathcal{Q}=10$, consisting of two concentric domain walls. Note that on the outer ring, the fractional vortices follow a $123\ldots$ ordering, while on the inner ring, the order is $132\ldots$. The reversal occurs because from the perspective of the outer ring, the regime outside is $s+ \ii s$ and the inner is $s-\ii s$, while for the inner ring, the regimes are the opposite. We can also see that the outer ring forces the inner ring to contract, which is most apparent studying the magnetic field.

\section{Conclusion}

In conclusion 
we demonstrated that, at the level of a microscopic Bogoliubov-de Gennes theory in three-band $s+\ii s$ superconductors, there are flux carrying  \skyr skyrmions.
The skyrmions can be viewed as stable bound states of spatially separated fractional vortices and circular domain walls.
These solutions were obtained by solving a fully self-consistent three-band Bogoliubov-de Gennes model coupled to a gauge field.
In the considered parameter regime, we find that skyrmions are slightly more energetically expensive than vortices per unit of magnetic flux. 
This suggests that in Ba$_{\rm 1-x}$K$_{\rm x}$Fe$_2$As$_2$, where the $s+\ii s$ state has been reported, the magnetic response under external field along the $c$-axis should be dominated by ordinary vortices. However, a sufficiently strong quench of a lattice of ordinary vortices may produce \skyr skyrmions. Recently more superconducting materials that break time reversal symmetry were discovered  \cite{ghosh2020recent}, in which skyrmion solutions can be further searched for.
When the superconducting bands are strongly coupled, we find that the magnetic signature of a skyrmion can manifest itself in a scanning SQUID experiment as a washed-out circular, or more generally, closed stripe of magnetic flux. 
Another possible route for detection of skyrmions is scanning tunneling spectroscopy. In fact, the skyrmion solutions that we presented exhibit clear signatures in the local density of states correlated with the position of fractional-flux vortices.

\section{Acknowledgments}
We thank Albert Samoilenka for insightful discussions. The work was supported by the Swedish Research Council Grants  2016-06122, 2018-03659.

\appendix

\section{Rescaling} \label{appendix: rescaling}

In dimensionful units, the Hamiltonian for our three-component superconductor, before the mean-field approximation, reads
\begin{equation}
\begin{aligned}
        H = & - t\sum_{\alpha\sigma}\sum_{<ij>}\exp{\ii e a A_{ij} / \hbar }c^\dagger_{i\sigma\alpha}c_{j\sigma\alpha} \\ & -\sum_{i\alpha\beta}V_{\alpha\beta} c_{i\uparrow\alpha}^\dagger c_{i\downarrow\alpha}^\dagger c_{i\downarrow\beta} c_{i\uparrow\beta},
\end{aligned}
\end{equation}
where $t$ is the nearest-neighbor hopping parameter, $e$ is the electronic charge and $A_{ij} = \frac{1}{a}\int_j ^i \Vec{A} \cdot \dd{\Vec{\ell}}$ accounts for interaction with the magnetic vector potential through Peierls substitution ($a$ is the lattice spacing). The energy associated with the magnetic field equals
\begin{equation}
    F_\magnetic = \frac{1}{2\mu_0^{\rm{2D}}} \sum_{\rm{plaquettes}} B_z^2 a^2,
\end{equation}
where $B_z = (A_{21} + A_{32} - A_{34} - A_{41}) / a$ (following the notation introduced in the main text), and $\mu_0^{\rm{2D}} = \mu_0 /L_z$ is the effective two-dimensional vacuum permeability for the system ($L_z$ is the perpendicular lengthscale associated with the magnetic field). 

Let us introduce a dimensionless system where we measure all energies in units of the hopping parameter $t$, and the planar coordinates $x$ and $y$ in units of the lattice spacing $a$ ($x = ax'$ and $y=ay'$). Let $H = t H'$ and $F_\magnetic = t F_\magnetic'$, where $H'$ and $F_\magnetic'$ are the rescaled energies. We introduce $V_{\alpha \beta} = t V_{\alpha \beta}'$ for the rescaled Hubbard Hamiltonian.
The rescaled magnetic energy equals
\begin{equation}
    F_\magnetic' =  \frac{1}{2} \sum_{\rm{plaquettes}} {B_z' } ^2,
\end{equation}
where $\Vec{B'} = \Vec{B} a/\sqrt{\mu_0^{\rm{2D}}t} = \Vec{\nabla'} \cross \Vec{A'}$ and $\Vec{A'} = \Vec{A} / \sqrt{\mu_0^{\rm{2D}}t}$. The Peierls phase reads
\begin{equation}
    \frac{e}{\hbar} aA_{ij} = q A_{ij}',
\end{equation}
where $q = \frac{e}{\hbar} a \sqrt{\mu_0^{\rm{2D}}t}$ and $A_{ij}' = \int_j^i \Vec{A'} \cdot \dd{\Vec{\ell'}}$ are dimensionless. 
Throughout the main text of the paper we use these rescaled coordinates, and for brevity we drop the prime-notation.

\section{Derivation of mean-field Hamiltonian}\label{appendix: modelDerivation}
The partition function associated with the Hamiltonian in \eqref{eq: full Hubbard hamiltonian} is
\begin{equation}
    Z = \int\mathscr{D}[ c^\dagger c] \ee^{-S( c^\dagger c)} ,
\end{equation}
where the action $S$ reads
\begin{equation}
    \begin{aligned}
        S &=\int_0^{1/k_\B T} \dd\tau\left[ \sum_{ij\sigma\alpha} c^\dagger_{i\sigma\alpha}(\delta_{ij}\partial_\tau + h_{ij\alpha}) c_{j\sigma\alpha} \right. \\ 
        &- \left. \sum_{i\alpha\beta}V_{\alpha\beta} c_{i\uparrow\alpha}^\dagger c_{i\downarrow\alpha}^\dagger c_{i\downarrow\beta} c_{i\uparrow\beta}\right],
    \end{aligned}
\end{equation}
where $h_{ij} = h_{ji}^*$ is some general quadratic term (such as nearest neighbor hopping).
The mean-field approximation is used to re-write and approximate the quartic interaction term, defined by the interaction strengths $V_{\alpha \beta}$. We assume that the associated interaction matrix $V$ is non-singular, resulting in three superconducting order parameters (for discussion on the singular case, see \secref{sec: ground state}).
Let $\rho_{i\alpha} = { c_{i\downarrow\alpha} c_{i\uparrow\alpha}}$, such that the quartic interaction reads $-\sum_{i \alpha \beta} V_{\alpha \beta} \rho_{i\alpha}^\dagger \rho_{i\beta}$. By performing the Hubbard-Stratonovich transformation \cite{Hubbard_partition_1959,altland2010condensed} in the Cooper channel, we have
\begin{align}
    &\exp{\int\dd\tau\sum_{i\alpha\beta}V_{\alpha\beta}\rho_{i\alpha}^\dagger\rho_{i\beta}}=\int\mathscr{D} \qty[\Delta^*\Delta]\\ &\exp{-\int\dd\tau\sum_{i\alpha\beta}\qty[(V^{-1})_{\alpha_\beta}\Delta_{i\alpha}^\dagger\Delta_{i\beta}+\Delta_{i\alpha}^*\rho_{i\alpha}+\Delta_{i\alpha}\rho_{i\alpha}^\dagger]}\nonumber,
\end{align}
where $\Delta_{i\alpha}$ is an auxiliary bosonic field
and $V^{-1}$ the inverse of the coupling matrix.
Hence, the partition function becomes 
\begin{equation}
    \begin{aligned}
        Z =& \int\mathscr{D}[ \Psi^\dagger \Psi]\mathscr{D}[\Delta^*\Delta]\\ & {\rm{exp}}\Bigg\lbrace-\int\dd\tau\Bigg[\sum_{ij\alpha} \Psi^\dagger_{i\alpha}(\delta_{ij}\partial_\tau+M_{ij\alpha}) \Psi_{j\alpha} \\ & +\sum_{i\alpha\beta}(V^{-1})_{\alpha\beta}\Delta_{i\alpha}^*\Delta_{i\beta}\Bigg]\Bigg\rbrace,
    \end{aligned}
\end{equation}
where we introduced the Nambu spinor notation with 
\begin{equation}
     \Psi_{i\alpha} = \mqty( c_{i\uparrow\alpha} \\  c^\dagger_{i\downarrow\alpha}) \qquad  \Psi_{i\alpha}^\dagger = \mqty( c_{i\uparrow\alpha} &  c^\dagger_{i\downarrow\alpha}),
\end{equation}
and the matrix elements
\begin{equation}\label{eq: bandHamiltonian}
        M_{ij\alpha}=\mqty( h_{ij} &  \delta_{ij}\Delta_{i\alpha} \\ \delta_{ij}\Delta_{i\alpha}^* & -h^*_{ij}).
\end{equation}
Let us Fourier transform the Nambu spinors from imaginary time to the Matsubara frequency space, i.e., 
\begin{equation}
     \Psi_{i\alpha}(\tau) = \sum_{m=-\infty}^\infty\Tilde{ \Psi}_{i\alpha m}\ee^{\ii\omega_m\tau},
\end{equation}
where $\omega_m = 2\pi k_\B T(m+1/2)$ are the Matsubara frequencies for fermions. By integrating out the fermionic degrees of freedom $\tilde{ \Psi}_{i\alpha m},\tilde{ \Psi}_{i\alpha m}^\dagger$ we obtain the partition function
\begin{align}
&Z=\int\mathscr{D}[\Delta^*\Delta] \\ &\exp{-\int\dd\tau\sum_{i\alpha\beta}(V^{-1})_{\alpha\beta}\Delta_{i\alpha}^*\Delta_{i\beta} + \sum_{\alpha m}\ln\det(\mathds{1} \omega_m+M_{\alpha})}.\nonumber
\end{align}
where
\begin{equation}
    M_\alpha = \mqty(h & \Delta_\alpha \\ \Delta_\alpha^\dagger & -h^*)
\end{equation}
is a $2N\times 2N$ matrix, with $\Delta_\alpha=\textrm{diag}(\Delta_{1 \alpha}, \ldots, \Delta_{N \alpha})$ ($N$ is the total number of sites). 
By assuming that the auxiliary fields are classical, that is, they have no dependence on $\tau$, we can perform the imaginary time integral, yielding the partition function
\begin{equation}
    Z = \int\mathscr{D}[\Delta^*\Delta]\ee^{-S^\prime(\Delta_\alpha^*,\Delta_\alpha)},
\end{equation}
with
\begin{equation}
    S^\prime=\frac{1}{k_\B T} \sum_{i\alpha\beta} (V^{-1})_{\alpha\beta}\Delta^*_{i\alpha}\Delta_{i\beta} - \sum_{\alpha m }\ln\det(\mathds{1} \omega_m+M_{\alpha}). 
\end{equation}
At this point, we make the further assumption that the field does not have thermal fluctuation, which means that the only field configuration which does contribute to the partition function is the one which minimizes the action $S^\prime$. Therefore we can define the free energy of the model as $F_H=k_\B TS^\prime$, obtaining, after summing up the Matsubara frequencies
\begin{equation}
    F_H= \sum_{i} \Vec{\Delta}_i^\dagger V^{-1}\Vec{\Delta}_i + k_\B T\sum_\alpha\Tr\ln f(M_\alpha) ,
\end{equation}
where $f(x)=\qty(\ee^{\beta x} + 1)^{-1}$ is the Fermi-Dirac distribution and $\Vec{\Delta}_i=\qty(\Delta_{i1},\Delta_{i2},\Delta_{i3})^\T$. The last step to perform is to find the field configuration which minimizes the free energy, namely
\begin{equation}\label{eq: Fminimization}
    \pdv{F_H}{\Delta_{i\alpha}^*} = V^{-1}\Vec{\Delta}_i + f(M_\alpha)_{i,i+N} = 0.
\end{equation}
Following the derivations in \cite{nagai2012efficient,covaci2010efficient}, it is possible to show that the matrix elements of $f(H_\alpha)$ correspond to thermal averages. Explicitly, we have
\begin{align}
    \langle c_{i\uparrow\alpha}^\dagger c_{j\uparrow\alpha} \rangle &= f(M_\alpha)_{ji} \label{eq: matrix element up avg},\\
    \langle c_{i\downarrow\alpha}^\dagger c_{j\downarrow\alpha} \rangle &= \delta_{ij}-f(M_\alpha)_{i+N,j+N}, \label{eq: matrix element down avg}\\
    \langle c_{i\uparrow\alpha} c_{j\downarrow\alpha}\rangle &= -f(M_\alpha)_{j,i+N} \label{eq: matrix element anomalous},
\end{align}
yielding the self-consistency equations
\begin{equation}\label{eq: selfconsistent}
    \Delta_{i\alpha}=\sum_{\beta}V_{\alpha\beta}\langle c_{i\uparrow\alpha} c_{i\downarrow\alpha}\rangle.
\end{equation}
The fields $\Delta_{i\alpha}$ represent the superconducting gaps in each band. We can now insert back the equations for the gaps $\Delta_{i\alpha}$ into the free energy, obtaining, up to a constant
\begin{equation}\label{eq: freeenergy}
  F_H= \sum_{i} \Vec{\Delta}_i^\dagger V^{-1}\Vec{\Delta}_i -k_\B T\Tr\ln\qty(\ee^{-\beta M}+1),
\end{equation}
where $M=M_1 \oplus M_2 \oplus M_3$.
From \eqref{eq: selfconsistent} and \eqref{eq: bandHamiltonian} we can see, following the decoupling in the Cooper channel, that the three bands are described by single band matrices $M_\alpha$, and are coupled through the self-consistency equations.

\section{Numerical method details}\label{appendix: numericalMethods}
In order to solve the mean-field Hamiltonian self-consistently, we need to evaluate thermal averages to compute the superconducting gaps in \eqref{eq: selfConsistency} and the currents in \eqref{eq: currentDensity}. The standard approach, would be to express these thermal averages in terms of the quasi-particle wavefunctions $\left(u_{\uparrow \alpha i}^{(n)}, v_{\downarrow \alpha i}^{(n)}\right)^\T$, that satisfy the equation
\begin{equation}
    E_{n \alpha} \begin{pmatrix}
        u_{\uparrow \alpha i}^{(n)} \\ v_{\downarrow \alpha i}^{(n)}
    \end{pmatrix}
    =
    \sum_j \begin{pmatrix}
        h_{ij} & \Delta_{\alpha i} \delta_{ij} \\
        \Delta^*_{\alpha i} \delta_{ij} & -h_{ij}^*
    \end{pmatrix}
    \begin{pmatrix}
        u_{\uparrow \alpha j}^{(n)} \\ v_{\downarrow \alpha j}^{(n)}
    \end{pmatrix},
\end{equation}
where $h_{ij}$ corresponds to the hopping terms in the Hamiltonian. These quasi-particle wavefunctions can be found by diagonalizing matrices $M_\alpha$, which are $2N \times 2N$ Hermitian matrices. However, finding all wavefunctions and corresponding eigenvalues is computationally expensive. Since we are interested only in local thermal averages, such as those for the superconducting gaps and the currents, it is computationally more efficient to express these thermal averages as matrix elements of $f(M_\alpha)$, where $f(x)$ is the Fermi-Dirac distribution function (see \eqref{eq: matrix element up avg}, \eqref{eq: matrix element down avg} and \eqref{eq: matrix element anomalous}). Calculating these specific matrix elements is more tractable than full diagonalization of the matrices. We employ the Chebyshev spectral expansion method \cite{covaci2010efficient,nagai2012efficient,weisse2006kernel}, where the Fermi-Dirac distribution is approximated by an expansion in Chebyshev polynomials. For all the simulations, we use 400 polynomials to approximate the distribution function. The number of polynomials needed depends on the smoothness of the Fermi-Dirac distribution, which, in turn, depends on the temperature. For the temperature used in all the simulations we performed, 400 polynomials are sufficient to ensure highly accurate results. 

Next, we need to describe the iterative method used to update the magnetic vector potential and the superconducting gaps. For the superconducting gap we use
\begin{equation}
    \Delta_{i \alpha}^{(t+1)} = m \Delta_{i \alpha}^{(t)} + (1-m) \sum_\beta V_{\alpha \beta} \expval{c_{\uparrow i \beta} c_{\downarrow i \beta}}^{(t)},
\end{equation}
where $m \in [0,1)$ is a memory parameter, which is necessary to ensure convergence in certain parameter regimes, where a negative eigenvalue the potential matrix $V$ of large magnitude can destabilize the iterative procedure. The vector potential is updated by performing one gradient descent step
\begin{equation}
    A_{ij}^{(t+1)} = A_{ij}^{(t)} - \gamma \left( \pdv{F_{\magnetic}}{A_{ij}}^{(t)} - J_{ij}^{(t)} \right),
\end{equation}
where $\gamma=0.1$ is a constant coefficient. These iterations are continued until a specific convergence criterion is met. We use the requirements that mean value of the relative change, over all sites (or links for the magnetic vector potential) should be less than $10^{-6}$. Efficient calculation of the local density of states can be performed using the Chebyshev expansion method. However, we compute the local density of states using \eqref{eq: LDOS formula}, by performing one diagonalization step after convergence, to obtain all information contained in the eigenstates.

\bibliography{references.bib}
\end{document}